\documentclass[12pt]{article}
\usepackage{amssymb}
\usepackage{color}
\usepackage[centertags]{amsmath}
\usepackage{amsfonts}
\usepackage{amsmath}
\usepackage{epsfig}         
\usepackage{graphicx}
\usepackage{bm}             
\usepackage{eucal}          
\usepackage{amsthm}

\def\#{{\sharp}}
\def\R{{\mathbb{R}}}

\def\Z{{\mathbb{Z}}}

\def\bq{\begin{equation}}
\def\eq{\end{equation}}
\def\bqa{\begin{eqnarray*}}
\def\eqa{\end{eqnarray*}}
\def\br{\begin{array}}
\def\er{\end{array}}
\def\lbeq(#1){\label{eqn:#1}}
\def\refeq(#1){{\rm (\ref{eqn:#1})}}
\def\lbth(#1){\label{th:#1}}
\def\refth(#1){{\rm Theorem \ref{th:#1}}}
\def\lbass(#1){\label{ass:#1}}
\def\refass(#1){{\rm Assumption \ref{ass:#1}}}
\def\lblm(#1){\label{lm:#1}}
\def\reflm(#1){{\rm Lemma \ref{lm:#1}}}
\def\lbdf(#1){\label{df:#1}}
\def\refdf(#1){{\rm Definition \ref{df:#1}}}
\def\lbprp(#1){\label{prp:#1}}
\def\refprp(#1){{\rm Proposition \ref{prp:#1}}}
\def\lbcor(#1){\label{cor:#1}}
\def\refcor(#1){{\rm Corollary \ref{cor:#1}}}
\def\lbrm(#1){\label{rm:#1}}
\def\refrm(#1){{\rm Remark \ref{rm:#1}}}
\def\bgdf{\begin{definition}}
\def\eddf{\end{definition}}
\def\bgth{\begin{theorem}}
\def\edth{\end{theorem}}
\def\bglm{\begin{lemma}}
\def\edlm{\end{lemma}}
\def\bgprp{\begin{proposition}}
\def\edprp{\end{proposition}}
\def\bgcor{\begin{corollary}}
\def\edcor{\end{corollary}}
\def\bgexm{\begin{example}}
\def\edexm{\end{example}}
\def\bgpf{\begin{proof}}
\def\edpf{\end{proof}}
\def\bgpbs{\begin{problems}}
\def\edpbs{\end{problems}}
\def\bgrm{\begin{remark}}
\def\edrm{\end{remark}}
\def\bgrms{\begin{remarks}}
\def\edrms{\end{remarks}}
\def\bgass{\begin{assumption}}
\def\edass{\end{assumption}}
\def\bgds{\begin{description}}
\def\edds{\end{description}}
\def\ben{\begin{enumerate}}
\def\een{\end{enumerate}}

\newcommand{\pa}{\partial}

\newcommand{\be}{\begin{equation}}
\newcommand{\ee}{\end{equation}}
\newcommand{\beq}{\begin{eqnarray}}
\newcommand{\eeq}{\end{eqnarray}}
\newcommand{\beqq}{\begin{eqnarray*}}
\newcommand{\eeqq}{\end{eqnarray*}}

\newtheorem{theorem}{Theorem}
\newtheorem{lemma}{Lemma}
\newtheorem{proposition}{Proposition}
\newtheorem{definition}{Definition}
\newtheorem{remark}{Remark}
\newtheorem{corollary}{Corollary}
\newtheorem{example}{Example}
\newtheorem{assumption}{Assumption}

\begin{document}
\title{\bf  {Negative order KdV equation with both solitons and kink wave solutions}}
\author{Zhijun Qiao$^1$ and Jibin Li$^2$}
\date{}
\maketitle

\begin{center}
{$^1$Department of Mathematics, The University of
Texas-Pan American\\
1201 W University Drive, Edinburg, TX 78539, U.S.\\
$^2$Department of Mathematics, Zhejiang Normal University, Jinhua, Zhejiang, 321004, China}\\
{\small Email: qiao@utpa.edu}
\end{center}

\begin{abstract}
  {In this paper, we report an interesting integrable equation
 that has both solitons and kink solutions.}
 The integrable equation we study is
$(\frac{-u_{xx}}{u})_{t}=2uu_{x}$, which  actually comes from the
negative KdV hierarchy and could be transformed to the Camassa-Holm equation through a gauge transform.
The Lax pair of the equation is derived to
guarantee its integrability, and furthermore
the equation is shown to have classical solitons, periodic soliton and
kink solutions.


\end{abstract}
 {PACS. 02.30.Ik 每 Integrable systems.
}

\section{Introduction}


   Soliton theory and integrable systems play an important
 role in the study of nonlinear water wave equations. They have many significant applications
 in fluid mechanics, nonlinear optics, classical and quantum fields theories etc.
 Particularly in recent years, more focuses have been pulled to
 integrable systems with non-smooth solitons, such as peakons, cuspons,
 since the study of the remarkable
 Camassa-Holm (CH) equation with peakon solutions \cite{CH}.
Henceforth, much progress have been made in the study of non-smooth solitons
for integrable equations [4-20, 22, 27-33].

  In this paper, we consider the following integrable equation
  \bq  (\frac{-u_{xx}}{u})_{t}=2uu_{x}, \lbeq(KdV) \eq
  which is actually the first member in the negative KdV hierarchy  \cite{Zhijun Qiao2}.
Equation \refeq(KdV) is proven equivalent to the Camassa-Holm (CH) equation: $m_t+m_xu+2mu_x=0,m=u-u_{xx}$ through a gauge transform (see Remark 1 in the paper).
Therefore, we find a simpler reduced form of the CH equation.
The Lax pair of the equation \refeq(KdV) is derived to
guarantee its integrability, and furthermore
the equation is shown to have classical solitons, periodic solitons and
kink solutions.

\section{Derivation of Equation \refeq(KdV) and Lax Representation}

  Let us consider the Schr\"{o}dinger-KdV spectral problem:
 \bq L\psi\equiv\psi_{xx}+v\psi=\lambda\psi, \lbeq(SKdV) \eq
 where $\lambda$ is an eigenvalue, $\psi$ is the eigenfunction corresponding to the eigenvalue $\lambda$,
 and $v$ is a potential function.
 One can easily get the following Lenard operator relation:
 \bq K\nabla\lambda=\lambda J\nabla\lambda, \lbeq(eigen) \eq
 where $\nabla\lambda\equiv\frac{\delta\lambda}{\delta v}=\psi^{2}$ is the functional gradient of
 the spectral problem \refeq(SKdV) with respect to $v$,
 $K=\frac{1}{4}\partial^{3}+\frac{1}{2}(v\partial+\partial v)$ and $J=\partial$ are two Hamiltonian operators as known in the literature
 \cite{AS1}.

 By setting $v=-\frac{u_{xx}}{u}$, we have the product form of operators $K$, ${\cal L}$, $L$ and their
 inverses
 \begin{center}
  $K=\frac{1}{4}u^{-2}\partial u^{2}\partial u^{2}\partial u^{-2}$,
  $K^{-1}=4u^{2}\partial^{-1}u^{-2}\partial^{-1}u^{-2}\partial^{-1}u^{2}$,\\

  ${\cal L}=\frac{1}{4}\partial^{-1}u^{-2}\partial u^{2}\partial u^{2}\partial u^{-2}$,
  ${\cal L}^{-1}=4u^{2}\partial^{-1}u^{-2}\partial^{-1}u^{-2}\partial^{-1}u^{2}\partial$, \label{Recinverse}\\

  $L=\partial^{2}+v=u^{-1}\partial u^{2}\partial u^{-1}$,
  $L^{-1}=u\partial^{-1}u^{-2}\partial^{-1}u$,\\
\end{center}
where ${\cal L}=J^{-1}K$ and its inverse ${\cal L^{-1}}=K^{-1}J$ are the recursion operators for the positive order and negative order KdV
hierarchy that we study below.

Now, according to the Lenard's operators $K$ and $J$, we construct the entire KdV
hierarchy, and then we show the
integrability of the hierarchy through solving a key operator equation.

Let $G_0\in Ker\ J=\{G\in C^{\infty}(\R)\ |\ JG=0\}$ and $G_{-1}\in
Ker\ K=\{G\in C^{\infty}(\R)\ |\ KG=0\}$. We define the Lenard's
sequence
\beq G_j=\left\{\begin{array}{cc}
 {\cal L}^j G_0, \ j\in \Z,\\
 {\cal L}^{j+1} G_{-1}, \ j\in \Z,
\end{array}
\right. \label{Gj} \eeq where ${\cal L}, {\cal L}^{-1}$ are defined
by Eq. (\ref{Recinverse}).
Therefore  we generate a hierarchy of nonlinear evolution equations (NLEEs):
\beq v_{t_k}=J G_k=K G_{k-1}, \ \forall k \in \Z, \label{mtk} \eeq
which is called the entire KdV hierarchy. We will see below that the positive order $(k\ge0)$ gives
the regular KdV hierarchy usually mentioned in the literature \cite{AS1}, while the
negative order $(k<0)$ produces some interesting equations gauge-equivalent to the Camassa-Holm equation \cite{CH}.
Apparently, this hierarchy possesses the bi-Hamiltonian
structure because of the Hamiltonian properties of $K, J$.
Let us now give special equations in the entire KdV hierarchy (\ref{mtk}).

\begin{itemize}

\item Choosing $G_0=2\in Ker \ J$ (therefore $G_1=u$) leads to
      the second  positive member of the hierarchy (\ref{mtk}):
 \beq
  v_{t_2}=\frac{1}{2}v_{xxx}+ \frac{3}{2}vv_{x}, \label{m+2}
 \eeq
which is exactly the well-known KdV equation. Here there is nothing new.
Therefore, the positive order $(k\ge0)$ in the hierarchy (\ref{mtk})
yields the regular KdV hierarchy usually studied in the literature \cite{AS1}.

\item  Now, let us find kernel elements $G_{-1} \in Ker \ K$ in order to get the negative member of the hierarchy (\ref{mtk}).
Due to the product form of $K$ and $K^{-1}$, $G_{-1}=K^{-1}0$ has the following three seed solutions:
\begin{center}
$G^{1}_{-1}=f(t_{n})u^{2}$,
$G^{2}_{-1}=g(t_{n})u^{2}\partial^{-1}u^{-2}$,
$G^{3}_{-1}=h(t_{n})u^{2}\partial^{-1}u^{-2}\partial^{-1}u^{-2}$,
\end{center}
where $f(t_{n}), g(t_{n}), h(t_{n})$ are three arbitrarily given functions with respect to the
time variables $t_{n}$, but independent of $x$.
They produce three iso-spectral ($\lambda_{t_k}=0$) negative order KdV hierarchies
of Eq. (\ref{mtk})
   \be v_{t_k}=J{\cal L}^{k+1}\cdot G^l_{-1}, \ l=1,2,3, \ \ k=-1,-2,....
   \label{6.61u} \ee
When $k=-1$, their representative equations are:
\beq
\left(-\frac{u_{xx}}{u}\right)_{t_{-1}} &=& 2f\left(t_n\right)uu_x,  \label{nkdvrep1}\\
\left(-\frac{u_{xx}}{u}\right)_{t_{-1}} &= & g\left(t_n\right)\left(2uu_x
          \pa^{-1}u^{-2}+1\right), \label{nkdvrep2}\\
\left(-\frac{u_{xx}}{u}\right)_{t_{-1}} &= & h\left(t_n\right)\left(2uu_x
    \pa^{-1}u^{-2}\pa^{-1} u^{-2}+
    \pa^{-1}u^{-2}\right). \label{nkdvrep3}
\eeq
\end{itemize}

\begin{remark}
Apparently, the first one is differential and simpler and
exactly recovers the equation \refeq(KdV) after setting $f\left(t_n\right)=1$, which we focus on in the current paper.
Actually, these three representative equations (\ref{nkdvrep1}),  (\ref{nkdvrep2}), and (\ref{nkdvrep3}) come from $v_{t_{-1}}=JG_{-1}=JK^{-1}0$ with $v=-\frac{u_{xx}}{u}$. Clearly, $v_{t_{-1}}=JK^{-1}0$ is equivalent to $KJ^{-1}v_{t}=0$ ($t=t_{-1}$), that is,
\beq
\Big(\frac{v_{txx}}{v_x}\Big)_x+4\Big(\frac{vv_{t}}{v_x}\Big)_x+2v_t=0. \label{nkdvrep}
\eeq
This equation is exactly the one studied by Fuchssteiner \cite{FB1} (see equations (7.1) and (7.22) there. Equations (7.1) in \cite{FB1} has a typo and should be same as (\ref{nkdvrep}). From \cite{FB1}, the Camassa-Holm (CH) equation  is gauge-equivalent to  equation (\ref{nkdvrep}) through some hodograph transformations (7.11) and (7.12) in \cite{FB1}. In our paper, through using $v=-\frac{u_{xx}}{u}$ we further reduce equation (\ref{nkdvrep}) to a more simple form (i.e. equation \refeq(KdV)):
\beq
\Big(-\frac{u_{xx}}{u}\Big)_t=2uu_x. \lbeq(nKdV)
\eeq
In other words, we found a very interesting fact that equation \refeq(nKdV) can be viewed as a reduction form of the CH equation due to the above gauge-equivalence. In next section, we will solve this form.
\end{remark}

 {In the paper \cite{H1}, the author dealt with equation $(\partial^2 +4v+2v_x\partial^{-1})v_t=0$ by using the positive KdV hierarchy approach, and
all soliton solutions were given implicitly. This equation could be transformed to $(-u_{xx}/u)_t=2uu_x$ through $v=-u_{xx}/u$, like we mentioned earlier in our paper, but, this is only one of three reductions. So, solutions of this equation can not give all solitons of our equation $(-u_{xx}/u)_t=2uu_x$. In our paper, we present all solitons and kink solutions in a explicit form. Also, there is the connection of the first negative KdV equation $(\partial^2 +4v+2v_x\partial^{-1})v_t=0$ with sine-Gordon \cite{HW1}. But, the equation $(-u_{xx}/u)_t=2uu_x$ we propose in the current paper is not equivalent to the sine-Gordon equation, because the sine-Gordon equation has only kink solution while our equation has both kink solutions and classical solitons.}

Of course, we may generate higher order nonlinear equations by selecting
different members in the hierarchy. In the following, we will see
that all equations in the KdV hierarchy (\ref{mtk}) are integrable.
Particularly, {\sf the above three equations  (\ref{nkdvrep1}), (\ref{nkdvrep2}), and (\ref{nkdvrep3})
are integrable}.

Let us return to the spectral problem \refeq(SKdV). Apparently, the Gateaux derivative
matrix  $L_{*}(\xi)$ of the spectral operator  $L$  in the direction
$\xi\in C^{\infty}(\R)$ at point $v$ is \be
L_{*}(\xi)\stackrel{\triangle}{=}\left.\frac{{\rm d}}{{\rm d}
\epsilon} \right|_{\epsilon=0}U(u+\epsilon\xi) = \xi \label{6.4.5} \ee which is obviously an
injective homomorphism, i.e. $U_*(\xi)=0\Leftrightarrow \xi=0 $.

   For any given $C^{\infty}$-function $G$, one may consider
    the following operator equation \cite{CAO1} with respect to
    $V=V(G)$
    \beq   [V, L]=L_{*}(K G)- L_{*}(J G)L. \label{VLCH}
    \eeq

  \begin{theorem}
For the spectral problem \refeq(SKdV) and an arbitrary
$C^{\infty}$-function $G$, the operator equation (\ref{VLCH})  has the
following solution
\beq
  V=-\frac{1}{4}G_x+\frac{1}{2}G\pa,
 \label{6.4.15}
\eeq where $\pa=\pa_x=\frac{\pa}{\pa x}$, and subscripts stand for the partial derivatives in
$x$.  \label{Th1}
\end{theorem}
{\bf Proof:} A direct substitution will complete the proof.


\begin{theorem}
Let $G_0\in {\rm Ker} \ J$, $G_{-1}\in {\rm Ker} \ K$, and let each
$G_j$ be given through the Lenard sequence (\ref{Gj}). Then,
\begin{enumerate}
\item
  each new vector field $X_k=J G_k, \ k\in \Z$ satisfies the
following commutator representation
\beq L_{*}(X_k)=[V_k, L], \
\forall k\in \Z; \label{UV1} \eeq
\item  the entire KdV hierarchy (\ref{mtk}), i.e.
\beq v_{t_k}= X_k=J G_k, \ \forall k \in \Z, \label{mtk1} \eeq
possesses the Lax representation \beq
L_{t_k}=[V_k, L],\ \forall k\in \Z, \label{XUV} \eeq
\end{enumerate}
where \beq V_k&=&\sum V(G_j)L^{(k-j-1)}, \ \ \sum\ = \
\left\{\begin{array}{ll}
\sum^{k-1}_{j=0}, & k>0,\\
0, & k=0,\\
-\sum^{-1}_{j=k}, & k<0,
\end{array}\right.
\eeq and $V(G_j)$ is given by Eq. (\ref{6.4.15}) with $G=G_j$.
\label{Th12}
\end{theorem}

{\bf Proof:}  Let us only prove the case for $k<0$. We have \beqq
[V_k, L]&=&-\sum_{j=k}^{-1}[V(G_j), L]L^{k-j-1}\\
&=& -\sum_{j=k}^{-1}(L_*(K G_j)-L_*(JG_j)L)L^{k-j-1}\\
&=& -\sum_{j=k}^{-1} L_*(K G_j)L^{(k-j-1}-L_*(KG_{j-1})L^{k-j}\\
&=& L_*(K G_{k-1})-L_*(K G_{-1})L^{k}\\
&=& L_*( K G_{k-1})=L_*( J G_k)\\
&=& L_*(X_k). \eeqq

  Noticing $L_{t_k}=L_*(v_{t_k})$, we have
\beqq L_{t_k}-[V_k, L]= L_*(v_{t_k}-X_k). \eeqq The
injectiveness of $L_*$ implies the second result holds.

So, the entire KdV hierarchy  (\ref{mtk}) has  Lax pair and all equations in
the hierarchy are therefore integrable. In particular, the KdV
equation (\ref{m+2}) has the Lax pair $L_{t_1}=[W_1,L]$ with $L=\pa^2+v$ and $W_1=\pa^3+\frac{3}{2}v\pa+\frac{3}{4}v_x$,
which was well-known in the literature \cite{AS1}. Interesting thing is that
the first member $(k=-1)$ in the negative order KdV hierarchy (\ref{6.61u}) has the standard Lax representation $L_{t_{-1}}=
[V_{-1}^l,L]$
with $V^l_{-1}=
\left(\frac{1}{4}G^l_{-1,x}-
\frac{1}{2}G^l_{-1}\pa\right)L^{-1}$,
$l=1,2,3,$ $L=\partial^{2}+v=u^{-1}\partial u^{2}\partial u^{-1}$, and
  $L^{-1}=u\partial^{-1}u^{-2}\partial^{-1}u$. In particular, the negative KdV equation (\ref{nkdvrep1}) possesses the following Lax form:
$L_{t_{-1}}=[V^1_{-1},L]$
with
\beqq
V^1_{-1}&=&\left(\frac{1}{4}G^1_{-1,x}-\frac{1}{2}G^1_{-1}\pa\right)L^{-1}\\
        &=& \Big(\frac{1}{2}uu_x-\frac{1}{2}u^2\pa\Big)L^{-1}\\
        &=& -\frac{1}{2}u\pa^{-1}u.
\eeqq
All of those negative members in the hierarchies (\ref{6.61u}) are integrable.

\section{All traveling wave solutions of \refeq(KdV)}

 Let us now consider the traveling wave solution of equation \refeq(KdV)
 through a generic setting $u(x,t)=U(x-ct)$, where $c$ is the wave speed. Let $\xi=x-ct$, then
 $u(x,t)=U(\xi)$. Substituting it into equation \refeq(KdV) yields
\bq c(\frac{U^{''}}{U})^{'}=2UU^{'}. \lbeq(KdV1) \eq Integrating it
once, we obtain the following standard cubic Hamiltonian system
for $c\neq 0:$ \bq \frac{dU}{d\xi}=y=\frac{\pa H}{\pa y},\ \ \
\frac{dy}{d\xi}=gU+\frac1c{U^3}=-\frac{\pa H}{\pa U}, \lbeq(KdV2)    \eq
where $g$ is
an integral constant, and the Hamiltonian function is\bq
H(U,y)=\frac12y^2-\frac12gU^2-\frac{1}{4c}U^4. \lbeq(KdVr5) \eq
When $gc\geq 0,$ (21) has only one equilibrium point $O(0,0)$.
When $gc<0$, (21) has three equilibrium points $O(0,0)$ and
$E_{1,2}(\pm\sqrt{|cg|},0)$. Write that
$$h_0=H(0,0)=0,\ \ \ \ h_1=H(\pm\sqrt{|cg|},0)=\frac14cg^2.$$

By qualitative analysis, we have the bifurcations of phase
portraits of (21) in the \\$(c,g)-$parametric plan shown in Fig.1
(1-1)-(1-6).

\begin{center}
\begin{tabular}{ccc}
\epsfxsize=4cm \epsfysize=4cm \epsffile{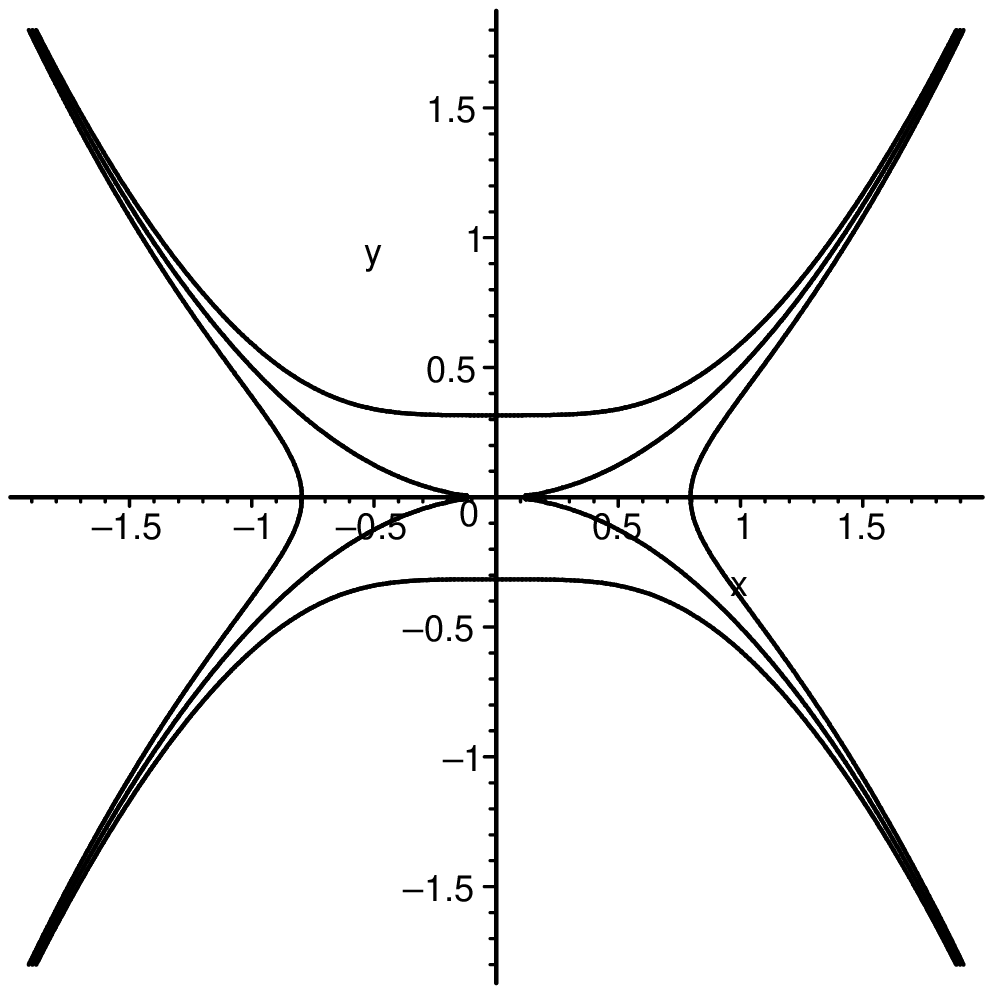} &
 \epsfxsize=4cm\epsffile{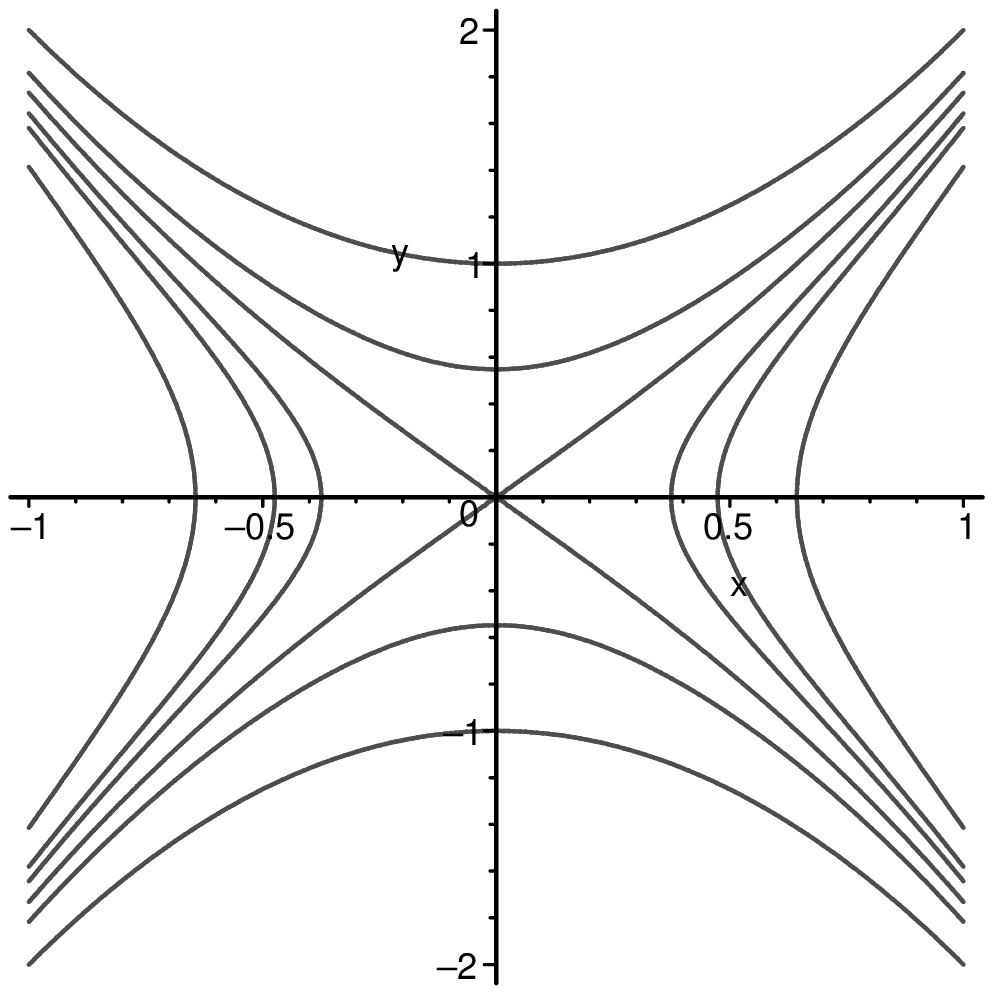} &
\epsfxsize=4cm \epsffile{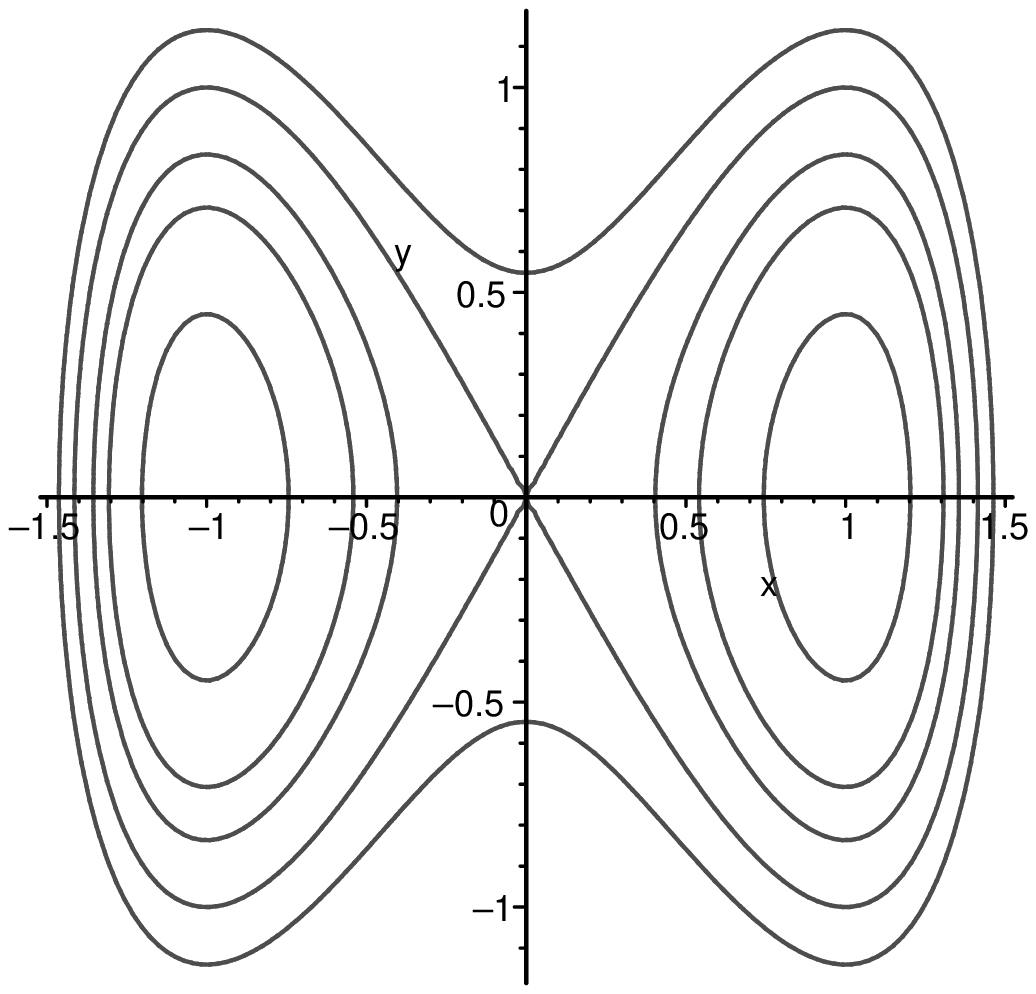}\\
(1-1) & (1-2) & (1-3)  \\
\end{tabular}
\end{center}

\begin{center}
\begin{tabular}{ccc}
\epsfxsize=4cm \epsfysize=4cm \epsffile{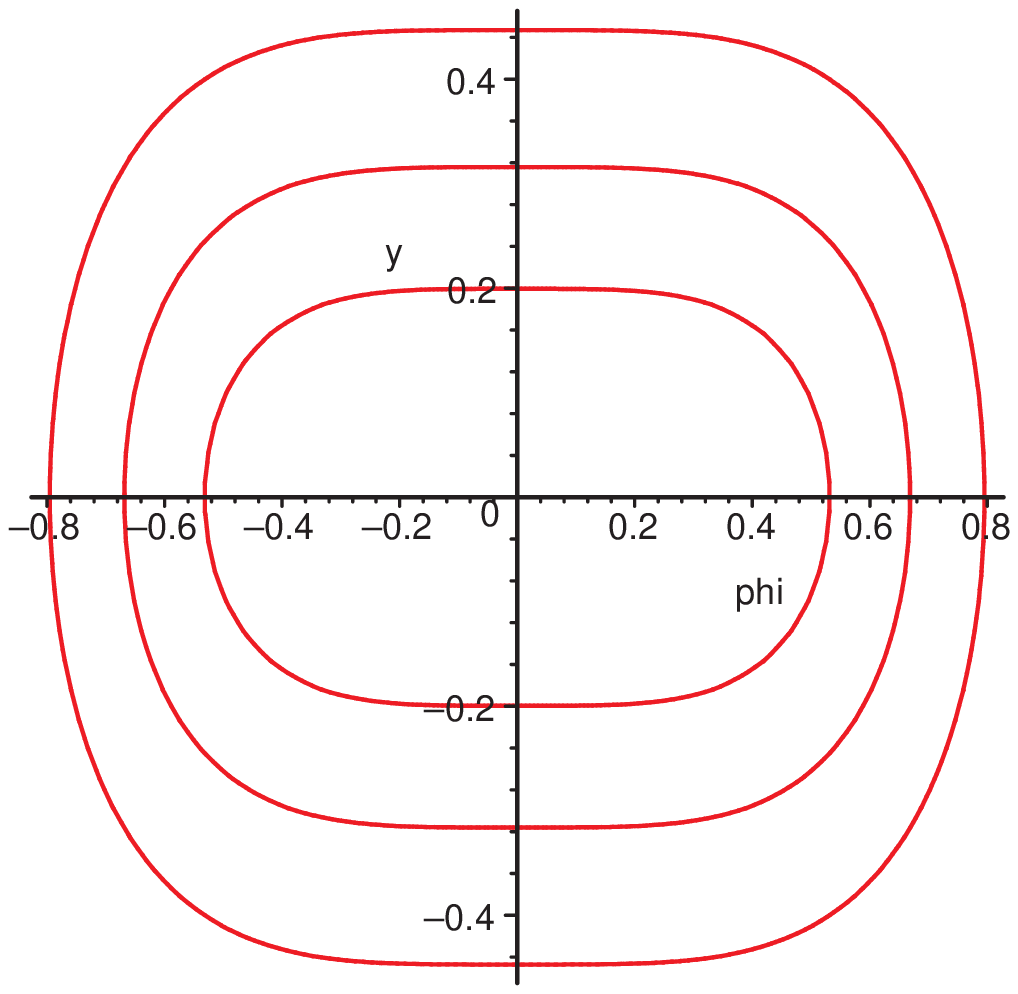} &
 \epsfxsize=4cm\epsffile{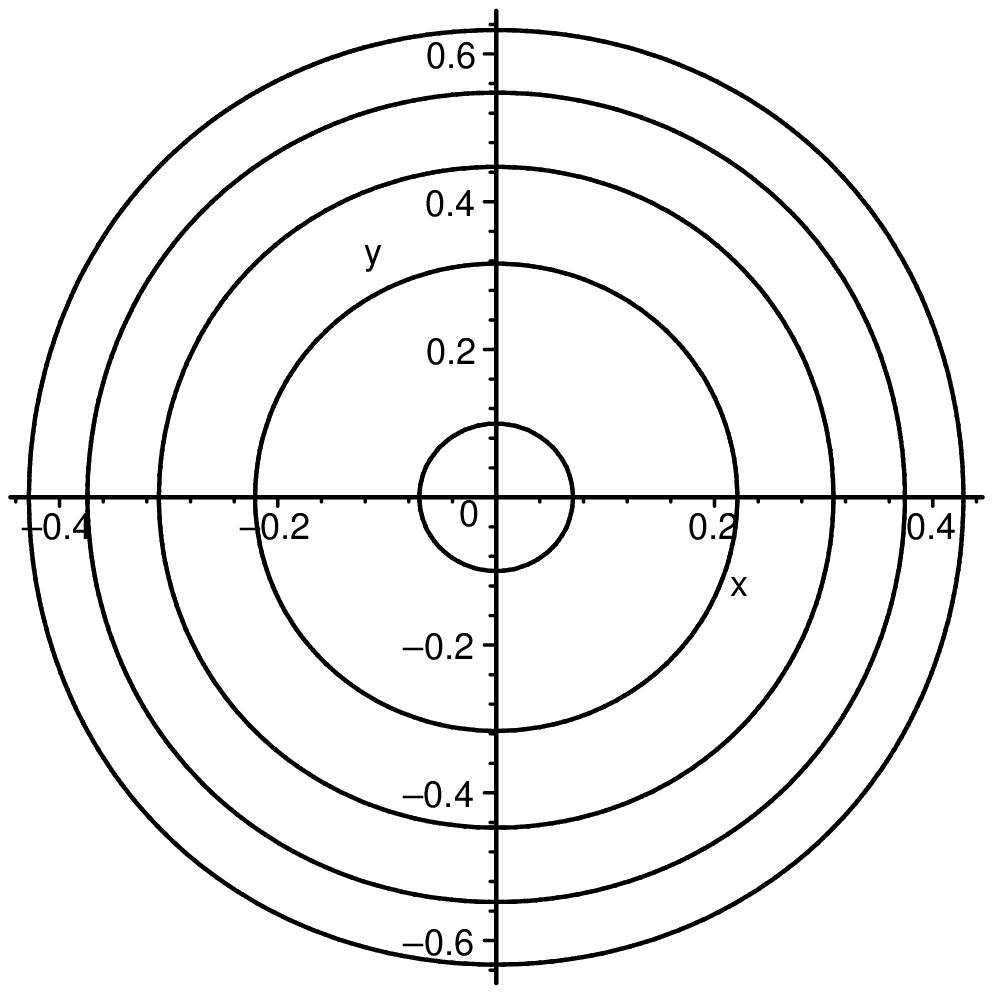} &
\epsfxsize=4cm \epsffile{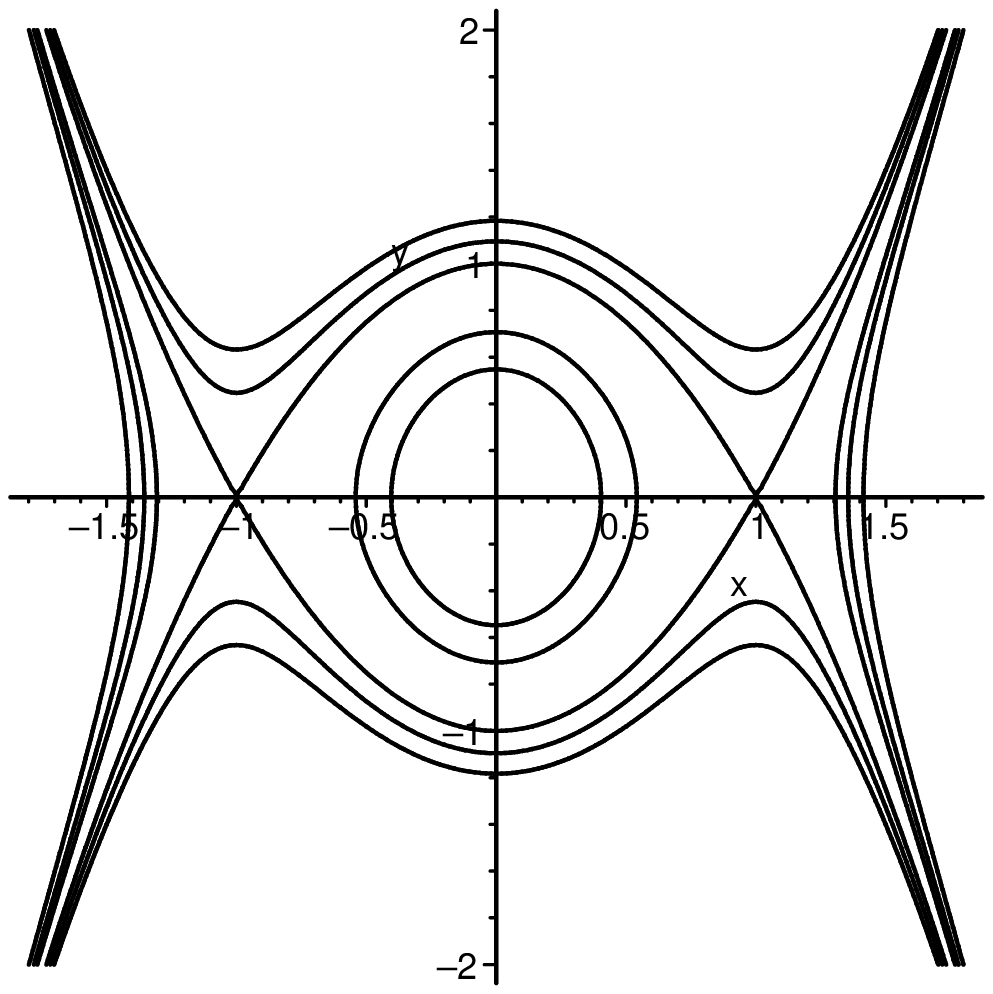}\\
(1-4) & (1-5) & (1-6)\\
\multicolumn{3}{c}{ Fig. 1   The change of phase
portraits of (21) in the $(c,g)-$parameter plan.} \\
\end{tabular}
\end{center}
{\footnotesize (1-1) $g=0, c>0.$ (1-2) $g>0, c>0.$ (1-3) $g>0,
c<0.$  (1-4) $g=0, c<0.$ (1-5) $g<0, c<0.$ (1-6) $g<0, c>0$.

\normalsize

Next, we present the exact traveling wave solutions of
(1) in an explicit form.

{\bf Case 1: $c>0, g=0$} (see Fig.1 (1-1)).

In this case, corresponding to 
the saddle point $O(0,0)$  \refeq(KdV2) reads as 
$y=\pm\frac{U^2}{\sqrt{2c}}$.  Using the first equation of (21)
and taking integration, we obtain
 \bq U(\xi)=\mp\frac{\sqrt{2c}}{\xi
+\xi_0}, \ \ \ \ \xi=x-ct,\lbeq(KdV3) \eq where $\xi_0$ is an
initial value of $\xi$. Clearly, when $\xi\rightarrow-\xi_0,
U(\xi)\rightarrow\infty.$ i.e., $U(\xi)$ is unbounded at
$\xi=\xi_0$. Thus, we have two unbounded breaking wave solutions
shown in Fig.2.

\begin{center}
\begin{tabular}{c}
\epsfxsize=5cm \epsfysize=5cm \epsffile{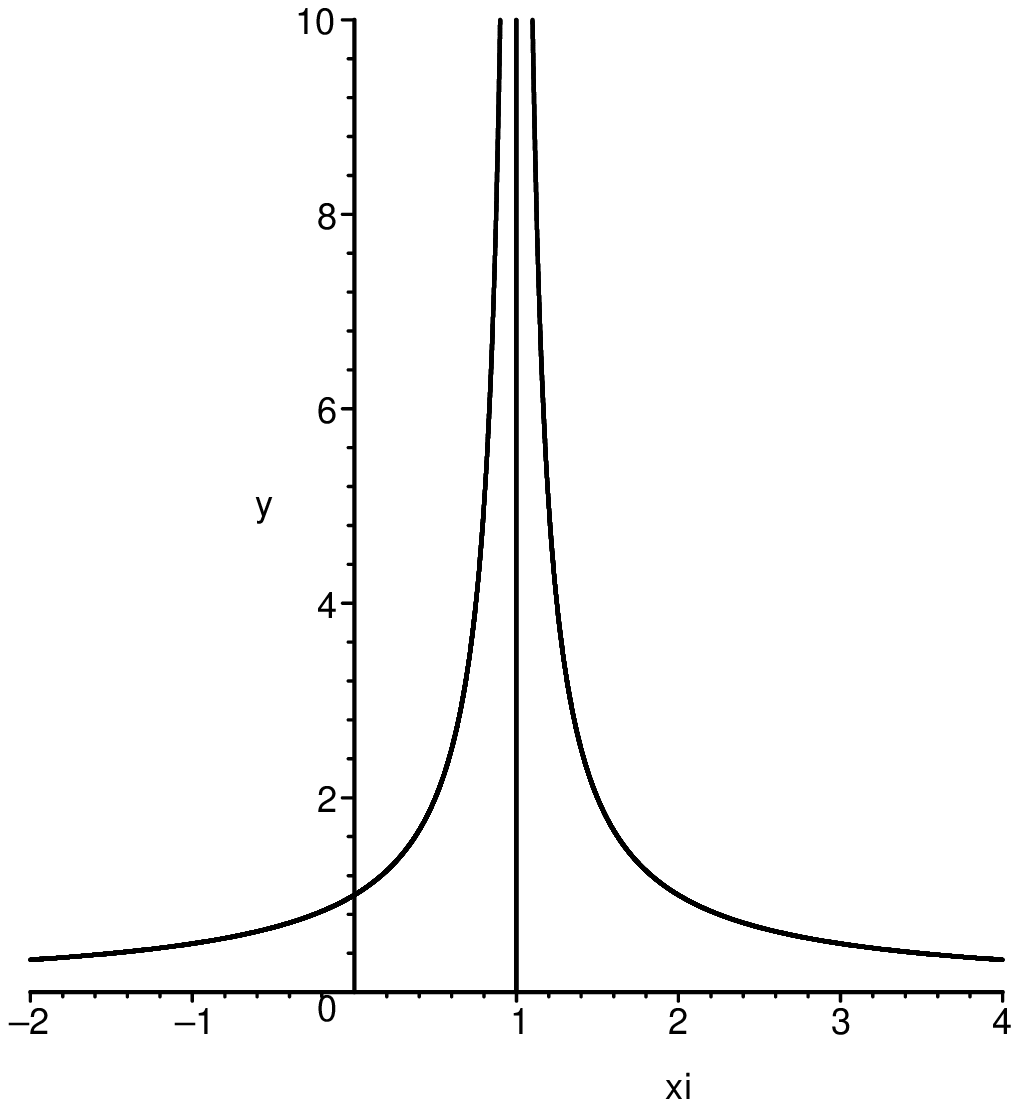}\\
\multicolumn{1}{c}{ Fig. 2   The profiles of the functions (23) where $\xi_0=-1$.} \\
\end{tabular}
\end{center}

{\bf Case 2: $c>0, g>0$} (see Fig.1 (1-2)).

Corresponding to 
the saddle
$O(0,0)$  \refeq(KdV2) reads as $y^2=gU^2+\frac{1}{2c}U^4$.
Using the first equation of (21) to take integration, we obtain
 \bq U(\xi)=\pm\frac{8Acg}{A^2e^{\sqrt{g}(\xi+\xi_0)}-8cge^{-\sqrt{g}(\xi+\xi_0)}},
\lbeq(KdV4) \eq where $A$ is an integrant constant.  When
$A^2=8cg$, the functions defined by (24) are discontinuous at
$\xi=-\xi_0.$ The profiles of (24) like Fig.2.

{\bf Case 3: $g>0, c<0$} (see Fig.1 (1-3)).

There exist three equilibrium points of (21) at $E_{1,2}$ and
$O(0,0)$. $O$ is a saddle point, $E_{1,2}$ are center points.

Corresponding to two homoclinic orbits, defined by $H(U,y)=0$, we
have the parametric representations
$$U(\xi)=\pm\sqrt{2|c|g}\textmd{sech}\sqrt{\frac{|c|g}{2}}\xi.\eqno(25)$$
They give two soliton solutions of (1).

For $h\in(\frac{1}{4}cg^2,0),$ corresponding to two families of
periodic orbits of (21), defined by $H(U,y)=h$, i.e.,
$y^2=\frac{1}{2|c|}(4|c|h+2g|c|U^2-U^4)=\frac{1}{2|c|}(r_1^2-U^2)(U^2-r_2^2)$,
where $r_1^2=g|c|+\sqrt{g^2c^2-4|c|h},
r_2^2=g|c|-\sqrt{g^2c^2-4|c|h},$  we obtain the parametric
representations of periodic wave solutions of (1) as follows:
$$U(\xi)=\pm r_1 \textmd{dn}\left(\frac{r_1}{\sqrt{2|c|}} \xi,
\frac{\sqrt{r_1^2-r_2^2}}{r_1}\right). \eqno(26)$$

For $h\in(0,\infty),$ corresponding to the family of periodic
orbits of (21), enclosing three equilibrium points defined by
$H(U,y)=h$, we have the following parametric representation of periodic wave
solutions of (1):
$$u(\xi)=r_1 \textmd{cn}\left(\sqrt{\frac{(r_1^2-r_2^2)}{2|c|}} \xi,
\frac{r_1}{\sqrt{r_1^2-r_2^2}}\right). \eqno(27)$$

{\bf Case 4: $g\leq0, c<0$} (see Fig.1 (1-4), (1-5)).

In this case, the origin $O(0,0)$ of (21) is an unique equilibrium
point, which is a center.  There exists a family of periodic orbits
of (21), enclosing the origin. (1) has the same parametric
representation of periodic wave solutions as (27).

{\bf Case 5: $g<0, c>0$} (see Fig.1 (1-6)).

In this case, there exist three equilibrium points of (21) at
$E_{1,2}$ and $O(0,0)$. $O$ is a center, $E_{1,2}$ are saddle
points. The heteroclinic orbits, defined by $H(u,y)=h_{1}$, have the
parametric representations
$$U(\xi)=\pm\sqrt{c|g|}\tanh\left(\frac{\xi}{\sqrt{2|g|}}\right),
\eqno(28)$$ which gives a kink wave solution and an anti-kink
wave solution of (1).

We see from (22) that $y^2=\frac{1}{2c}(4ch+2cgU^2+U^4.$ For
$h\in(0,h_1),$ it can be written as $y^2=\frac{1}{2c}(
(z_1^2-U2)(z_2^2-U^2), $ where
 $z_1^2=|g|c+\sqrt{g^2c^2-4ch},
z_2^2=|g|c-\sqrt{g^2c^2-4ch},$ Thus, the family of periodic orbits,
defined by $H(u,y)=h$,  has the parametric representation
$$u(\xi)=Z_2 \textmd{sn}\left(\frac{z_1\xi}{\sqrt{2c}},\frac{z_2}{z_1}\right), \eqno(29)$$
which gives a family of periodic wave solutions of (1).

\section{Conclusions}
In this paper, we reported an interesting property of integrable
system: solitons and kink solutions can occur in the
same integrable equation, and those solutions are given explicitly.
Within our knowledge, this is probably the first integrable example possessing such property.
We found this equation in the negative
order KdV hierarchy, which is gauge-equivalent to the CH equation.
Since equation \refeq(KdV) has the Lax pair, we may try to get
$r$-matrix structure of the constrained system of Lax equations,
and parametric and algebro-geometric solutions \cite{Q1}, but that
is beyond the scope of this paper. The symmetry of equations
(8), (9), and (10) were already discussed in \cite{Lou1}.
Recently, a two-fold  integrable  hierarchy associated with the KdV equation
was given in \cite{AK1}.  About other negative order integrable
hierarchies, such as the AKNS, the Kaup-Newell, the Harry-Dym , the Toda etc, one may see the literature \cite{Zhijun Qiao2}.

\section*{Acknowledgment}
  The Qiao's work is partially supported by the U. S. Army
  Research Office under contract/grant number W911NF-08-1-0511 and by the Texas Norman Hackerman Advanced Research Program under Grant No. 003599-0001-2009.

\end{document}